# Cinemática de un Fluido Ideal en un Universo Anisótropo Axisimétrico Espacialmente Plano

**López Ericsson[1]; Llerena Mario[1]; Aldás Franklin[1]**

[1]*Escuela Politécnica Nacional, Grupo de Relatividad General y Cosmología, Observatorio Astronómico de Quito*

**Resumen:** El Modelo Cosmológico Estándar asume que el Universo es, en promedio, homogéneo e isótropo para distancias por sobre los $10^9$ pc (z > 1), pero este principio se ha puesto en duda a partir de los resultados obtenidos del estudio de la Radiación Cósmica de Fondo, según los cuales, dicha radiación presenta anomalías anisótropas que no son explicadas desde el Modelo Estándar, como, por ejemplo, las fluctuaciones de temperatura del orden de $10^{-5}$ K o el alineamiento de momentos polares. Estas anomalías podrían explicarse a través de modelos cosmológicos anisótropos. En este trabajo proponemos una transformación a coordenadas esféricas considerando distintos factores de escala temporales en los ejes cartesianos, a partir de la cual se obtiene una métrica reducible a la métrica de Friedmann-Lemaître-Robertson-Walker con geometría espacial plana. Para el desarrollo del modelo consideramos el caso axisimétrico y analizamos el comportamiento cinemático de un fluido ideal en reposo.

**Palabras Clave:** Cosmología, anisotropía, expansión, rotación, *shear*.

## Kinematics of an Ideal Fluid into a Spatially Flat Anisotropic Axisymmetric Universe

**Abstract:** The Standard Cosmological Model assumes that the Universe is, on average, homogeneous and isotropic for large scales (z>1), but this principle has been questioned from the results about Cosmic Microwave Background. This radiation has anomalies that are not explained from the Standard Model, such as temperature fluctuations in the order of $10^{-5}$K or aligning polar moments. These anomalies could be explained by anisotropic cosmological models. We propose a transformation to spherical coordinates considering different temporal scale factors in the Cartesian axes, from which a reducible to flat spatial Friedmann-Lemaitre-Robertson-Walker metric is obtained. In the model, we consider the axisymmetric case and analyze the cinematic behavior of an ideal fluid at rest.

## 1. INTRODUCCIÓN

El Modelo Cosmológico Estándar asume que el Universo a grandes escalas está descrito, dentro del contexto de la Relatividad General, por la métrica de Friedmann-Lemaître-Robertson-Walker (FLRW) espacialmente plana, la cual es isótropa y homogénea. La métrica FLRW tiene la forma que se muestra en (1) donde R(t) es un factor de escala temporal.

$$ds^2 = dt \otimes dt - R^2(t)\left(dr' \otimes dr' + r'^2 d\Omega \otimes d\Omega\right) \quad (1)$$

Esta métrica cumple con el Principio Cosmológico (homogeneidad e isotropía), el cual se ha puesto en duda dados los resultados de misiones dedicadas al estudio de la Radiación Cósmica de Fondo (CMB).

Por ejemplo, el satélite *COsmic Background Explorer* (COBE) confirmó que la CMB mostraba un espectro de cuerpo negro con una temperatura de 2.736±0.010K con un nivel de confianza del 95 % pero se determinó que existían pequeñas fluctuaciones en la temperatura del orden de $\Delta T/T = 10^{-5}$ (Fernández, 2014). Posteriormente, con la misión *Wilkinson Microwave Anisotropy Probe* (WMAP), que operó desde el 2001 hasta el 2010, se hallaron anomalías para ángulos grandes en el espectro de potencias de las fluctuaciones de temperatura como el alineamiento entre el cuadrupolo y el octopolo (Akarsu et al., 2014; Fernández, 2014; Pavan y Pankaj, 2011; Russell et al., 2014), y el bajo valor del momento cuadrupolar (Russell et al., 2014). Finalmente, la misión Planck corroboró que las anomalías son características de la radiación y no son errores sistemáticos (Russell et al., 2014).

Otras anomalías halladas incluyen una región fría no gaussiana que no se esperaba dentro del modelo isótropo y la asimetría a gran escala entre los hemisferios del mapa de fluctuaciones (Fernández, 2014; Russell et al., 2014).

Existen otros resultados observacionales no vinculados con la CMB que podrían confirmar la violación del principio de isotropía, por ejemplo, la presencia de polarización en radio

ericsson.lopez@epn.edu.ec

de radio galaxias, la polarización en el visible de fuentes cuasi-estelares y el flujo de polarización en radio que sugieren direcciones preferenciales (Pankaj, n.d.; Pavan y Pankaj, 2011).

Toda esta evidencia observacional pone en duda la no existencia de direcciones preferenciales en el Universo, es decir, el Modelo Cosmológico Estándar podría estar basado en un principio de isotropía que no se cumple y por lo tanto, sus predicciones podrían estar, al menos, ligeramente erróneas.

Otro de los indicios para pensar en un Universo anisótropo es que, aunque hoy el Universo sea altamente isótropo, en épocas tempranas pudo no serlo y las anisotropías pudieron disminuir con el paso de su evolución (Cáceres, n.d.).

Para introducir en el Modelo Cosmológico las anomalías presentadas anteriormente, se puede pensar en introducir fuentes de materia anisótropas en el Modelo Estándar pero, por otro lado, y de manera más general, es posible resolver el problema al incluir las anisotropías como propiedades del espacio-tiempo e incluirlas en el tensor métrico, el cual es esencial para la descripción de la geometría del universo. Esto último es lo que se considera en los Modelos Bianchi, los cuales describen universos homogéneos pero anisótropos (Jacobs, 1968).

Existen varios tipos de Modelos Bianchi, pero un modelo Bianchi tipo I es el que se propone en esta contribución, principalmente porque su geometría espacial es plana, lo cual estaría acorde a las observaciones dado que los resultados de WMAP combinados con los resultados de las pruebas con supernovas tipo Ia sugieren que el Universo es espacialmente plano (Amanullah et al., 2010; Pradhan et al. 2015).

En coordenadas cartesianas, un modelo Bianchi I se desarrolla a partir de una métrica de la forma

$$ds^2 = dt \otimes dt - a^2(t) dx \otimes dx - b^2(t) dy \otimes dy - c^2(t) dz \otimes dz \quad (2)$$

donde a(t), b(t), c(t) son factores de escala que no necesariamente son iguales. La métrica en (2) es reducible a la métrica FLRW con geometría espacial plana.

Los modelos Bianchi I se han desarrollado en coordenadas cartesianas y se han estudiado algunos casos de contenido de materia (Russell et al., 2014) pero, en un espacio-tiempo axisimétrico, se podría considerar que las coordenadas naturales son las esféricas, además de que la métrica FLRW, base del Modelo Estándar, se la representa en estas coordenadas. Por otro lado, usar este nuevo sistema de coordenadas facilitaría la noción de movimiento radial.

Por otro lado, el estudio de la cinemática de un fluido ideal en un modelo cosmológico permite establecer la evolución temporal de un universo. Gracias a estudios de este tipo se ha podido caracterizar la expansión acelerada del Universo.

En este trabajo se plantea el desarrollo de las implicaciones en la cinemática de un fluido ideal en un modelo de Bianchi I axisimétrico en coordenadas esféricas. Se utilizará la siguiente notación: los índices con letras griegas van de 0 a 3 mientras que los índices con letras latinas van de 1 a 3.

## 2. PLANTEAMIENTO DE LA MÉTRICA

Para incluir la anisotropía en la parte espacial de la métrica se puede pensar en una transformación de coordenadas donde el factor a(t) no sea el mismo para cada coordenada cartesiana, es decir, una trasformación como la indicada en (3)

$$\begin{aligned} x &= r'a(t)\cos\theta\sin\phi \\ y &= r'b(t)\sin\theta\sin\phi \\ z &= r'c(t)\cos\phi \end{aligned} \quad (3)$$

con r'$\geq 0$ (distancia radial comóvil), $0 \leq \theta \leq 2\pi$ y $0 \leq \varphi \leq \pi$, donde, en principio, se cumple $a(t) \neq b(t) \neq c(t)$.

Dada la transformación de coordenadas, en general, las componentes $g_{\mu\nu}$ del tensor métrico corresponden a la expresión en (4)

$$g_{\mu\nu} = a_\mu^\lambda a_\nu^\lambda \quad (4)$$

donde se omite el símbolo de sumatoria en $\lambda$ y los elementos ($a^\mu{}_\nu$) de la matriz de cambio de coordenadas están dados por $a^\mu{}_\nu = \partial x^\mu / \partial y^\nu$.

Para el cambio de coordenadas propuesto en (3), se tiene que las componentes $g_{ij}$ no nulas del tensor métrico asociado a dicha transformación son las indicadas en (5).

$$\begin{aligned} g_{11} &= (a^2\cos^2\theta + b^2\sin^2\theta)\sin^2\phi + c^2\cos^2\phi \\ g_{22} &= r'^2(a^2\sin^2\theta + b^2\cos^2\theta)\sin^2\phi \\ g_{33} &= r'^2(a^2\cos^2\theta + b^2\sin^2\theta)\cos^2\phi + r'^2c^2\sin^2\phi \\ g_{12} &= -r'(a^2 - b^2)\cos\theta\sin\theta\sin^2\phi \\ g_{31} &= r'(a^2\cos^2\theta + b^2\sin^2\theta - c^2)\sin\phi\cos\phi \\ g_{23} &= -r'^2(a^2 - b^2)\cos\theta\sin\theta\sin\phi\cos\phi \end{aligned} \quad (5)$$

Al igual que en la métrica FLRW, se considerará que la métrica tiene la forma

$$ds^2 = dt \otimes dt - g_{ij} dx^i \otimes dx^j \quad (6)$$

y si, adicionalmente, en (6) se considera una variedad con simetría axial con el eje z coincidiendo con el eje de simetría, es decir, se considera a(t) = b(t), la métrica se reduce a la mostrada en (7)

$$\begin{aligned} ds^2 &= dt \otimes dt - (a^2\sin^2\phi + c^2\cos^2\phi)\,dr' \otimes dr' \\ &\quad - r'^2 a^2\sin^2\phi\,d\theta \otimes d\theta \\ &\quad - r'^2(a^2\cos^2\phi + c^2\sin^2\phi)\,d\phi \otimes d\phi \\ &\quad - 2r'(a^2 - c^2)\sin\phi\cos\phi\,dr' \otimes d\phi \end{aligned} \quad (7)$$

la cual es una métrica no diagonal en coordenadas esféricas que describe a un espacio-tiempo axisimétrico.

Se puede notar que en (7), el elemento de línea $d\Omega \otimes d\Omega$ en las hipersuperficies a r' = 1 y t constante es el indicado en (8)

$$d\Omega \otimes d\Omega = a^2 \sin^2\phi \, d\theta \otimes d\theta + (a^2 \cos^2\phi + c^2 \sin^2\phi) \, d\phi \otimes d\phi \tag{8}$$

y no corresponde al elemento de línea sobre un esfera de radio r' = 1, con lo cual, la variedad no tiene simetría esférica, salvo en el caso donde a(t) = c(t) que corresponde a la métrica FLRW.

### 3. ECUACIONES DE CAMPO

Las componentes G'$_{\mu\nu}$ del tensor de Einstein con constante cosmológica $\Lambda$ distinta de cero se muestran en (9).

$$G'_{\mu\nu} = R_{\mu\nu} - \frac{1}{2}g_{\mu\nu}R + g_{\mu\nu}\Lambda \tag{9}$$

Para establecer las componentes de los tensores de Riemann R$_{\mu\nu}$, de Ricci R, de Einstein con constante cosmológica $\Lambda$ y establecer el valor del escalar de curvatura se utilizó el paquete SageManifolds (Gourgoulhon et al., 2015) de SageMath (Stein et al., 2015).

Sea el cuadrivector velocidad de un fluido ideal en reposo u$_\mu$=∂/∂t y sean las componentes T$_{\mu\nu}$ del tensor energía-momento de dicho fluido las mostradas en (10)

$$T_{\mu\nu} = (\rho + p)(u_\mu u_\nu) - p g_{\mu\nu} \tag{10}$$

donde $\rho$ es la densidad de materia-energía y p la presión (Tanto la densidad de materia-energía como la presión son funciones sólo del tiempo). Para la métrica planteada en (7) se tiene que dicho tensor corresponde a (11).

$$\begin{aligned}T =\ & \rho \, dt \otimes dt + p(c^2\cos^2\phi + a^2\sin^2\phi)\, dr' \otimes dr' \\ & + pr'^2 a^2 \sin^2\phi \, d\theta \otimes d\theta \\ & + r'^2 p (a^2\cos^2\phi + c^2\sin^2\phi)\, d\phi \otimes d\phi \\ & + 2r'p(a^2 - c^2)\cos\phi\sin\phi \, dr' \otimes d\phi\end{aligned} \tag{11}$$

Planteado el tensor energía-momento se pueden establecer las ecuaciones de campo con fuentes de campo gravitatorio que son las indicadas en (12) con $\kappa$ una constante.

$$G'_{\mu\nu} = \kappa T_{\mu\nu} \tag{12}$$

A partir de (12), el sistema de ecuaciones a resolver y que describen al campo gravitatorio con fuentes de materia ideal en reposo en un universo axisimétrico es

$$\frac{a^2 c \Lambda + c\dot{a}^2 + 2a\dot{a}\dot{c}}{a^2 c} = k\rho \tag{13}$$

$$\frac{c^3\dot{a}^2 - a^3\dot{a}\dot{c} - a^4\ddot{c} - (a^4c - a^2c^3)\Lambda - (a^3c - 2ac^3)\ddot{a}}{a^2 c} = kp(a^2 - c^2) \tag{14}$$

$$\frac{a^2 c\Lambda + ac\ddot{a} + a\dot{a}\dot{c} + a^2\ddot{c}}{c} = -kpa^2 \tag{15}$$

que son las correspondientes a las ecuaciones de Friedmann en el caso axisimétrico que se plantea.

Además, dado que se debe cumplir con la ley de conservación $\nabla_\mu T^{\mu\nu} = 0$, se tiene que la densidad de materia y la presión evolucionan temporalmente cumpliendo con la Ecuación (16).

$$\dot{\rho} + (p + \rho)\left(2\frac{\dot{a}}{a} + \frac{\dot{c}}{c}\right) = 0 \tag{16}$$

### 4. SOLUCIÓN ECUACIONES DE FRIEDMANN AXISIMÉTRICAS

Para resolver el sistema de ecuaciones compuesto por (13), (14) y (15), más la ley de conservación (16), se consideran tres casos: el vacío, la época de dominio de polvo y la época de dominio de la radiación.

*4.1. Vacío con constante cosmológica*

En el vacío se tiene que T$_{\mu\nu}$ = 0. Para resolver las ecuaciones de campo con $\Lambda \neq 0$, y conocida la solución en FLRW, se propone una solución donde los factores de escala a(t) y c(t) sean proporcionales a $e^{\alpha t}$, donde $\alpha = \alpha(\Lambda)$.
Es decir, se proponen soluciones como en (17)

$$a(t) = K_a e^{\alpha t} \qquad c(t) = K_c e^{\alpha t} \tag{17}$$

con K$_a$ y K$_c$ constantes. Se puede comprobar que estas funciones son soluciones para las ecuaciones de campo si adicionalmente se cumple que $\alpha = \sqrt{-\Lambda/3}$.
Se puede notar que en el caso $\Lambda$=0, los factores de escala son constantes.

*4.2. Fluido ideal barotrópico*

Para las siguientes soluciones se considera un fluido ideal barotrópico que cumple con la Ecuación (18) como ecuación de estado.

$$p = \omega\rho, \qquad 0 \leq \omega \leq 1 \tag{18}$$

*4.2.1. Época dominada por Polvo*

Si se considera que el universo está dominado por polvo, es decir, partículas no interactuantes, se tiene que p=0.

Considerando Λ = 0 y conocido el resultado para FLRW, se propone una solución como en (19)

$$a(t) = K_a t^{2/3} \qquad c(t) = K_c t^{2/3} \qquad (19)$$

donde $K_a$ y $K_c$ son constantes. Se puede comprobar que estas funciones resuelven las ecuaciones de campo si adicionalmente, por la ley de conservación, se tiene que (20) también se satisface.

$$\rho(t) = \frac{4}{3\kappa} t^{-2} \qquad (20)$$

*4.2.2. Época dominada por Radiación*

Si se considera que el universo está dominado por radiación que cumple con la ecuación de estado p = ρ/3 y considerando que Λ = 0, se propone una solución del tipo mostrado en (21)

$$a(t) = K_a t^{1/2} \qquad c(t) = K_c t^{1/2} \qquad (21)$$

Con esta consideración y adicionalmente con (22)

$$\rho(t) = \frac{3}{4\kappa} t^{-2} \qquad (22)$$

se resuelven las ecuaciones de campo.

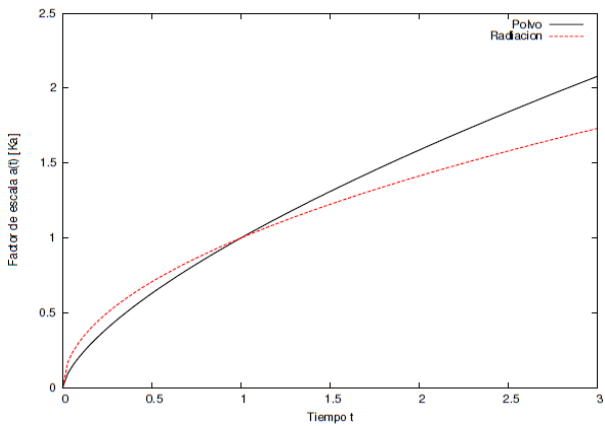

**Figura 1**. Evolución temporal del factor de escala a(t)

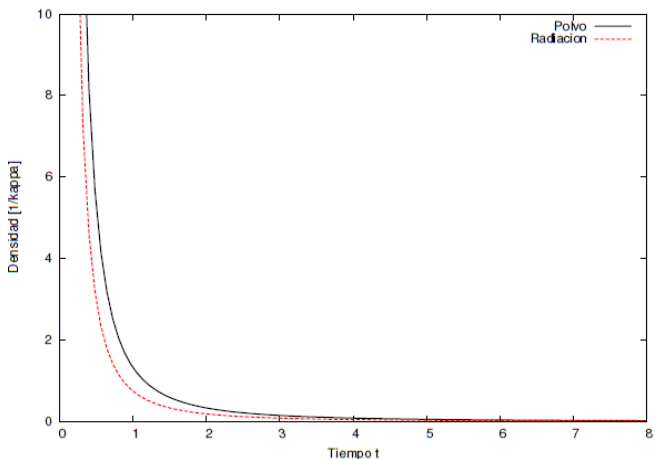

**Figura 2**. Evolución temporal de la densidad de materia-energía

En la Figura 1 se encuentra la evolución temporal del factor de escala a(t) en unidades de $K_a$ para los casos de dominio de polvo y radiación. Dado que, en principio, $K_a$ y $K_c$ son diferentes, la evolución temporal de a(t) y c(t) no debe ser la misma pero debe ser proporcional.

Se puede notar que, al igual que FLRW, el factor de escala en la época dominada por radiación crece más rápido en etapas iniciales del universo comparado con la época de dominio de polvo, pero este comportamiento se revierte para tiempos posteriores.

En la Figura 2 se encuentra la evolución temporal de la densidad de materia-energía ρ(t) en unidades de $\kappa^{-1}$ para los casos de dominio de polvo y radiación, la cual coincide con el resultado con FLRW, es decir, para tiempos lo suficientemente largos, la densidad de materia tiende a cero en ambos casos analizados. Así mismo, la densidad de radiación es siempre menor que la densidad de polvo en el universo que describimos.

Conocidas las tres soluciones anteriormente expuestas, podemos analizar el comportamiento cinemático de un fluido ideal en reposo en un universo axisimétrico espacialmente plano en los tres casos propuestos.

## 5. EXPANSIÓN DE UN FLUIDO IDEAL

Analizamos el movimiento de expansión de un fluido ideal en reposo con cuadrivelocidad $u_\mu = (1, 0, 0, 0)$. El escalar de expansión Θ, que mide la tasa a la cual la escala del universo va cambiando (incrementándose o decreciendo), está dado por (23).

$$\Theta = \frac{1}{3} u^\mu_{;\mu} \qquad (23)$$

donde, en (24) se muestra la expresión correspondiente a la derivada covariante del cuadrivector velocidad.

$$u^\mu_{;\nu} = \frac{du^\mu}{dx^\nu} + \Gamma^\mu_{\alpha\nu} u^\alpha \qquad (24)$$

Para la métrica en (7), se tiene que, si definimos a $H_a = \dot{a}/a$ y $H_c = \dot{c}/c$ como los parámetros de Hubble asociados a a(t) y c(t), respectivamente, entonces, el escalar de expansión, en función de los parámetros definidos, es el expuesto en (25).

$$\Theta = \frac{1}{3}(H_c + 2H_a) \qquad (25)$$

Para la métrica axisimétrica en el vacío con constante cosmológica no nula se tiene que $H_a = H_c$ y por lo tanto, $\Theta = \sqrt{-\Lambda/3}$, que corresponde al parámetro de Hubble en la métrica FLRW. Es decir, para el universo vacío, la expansión es constante y depende de la constante cosmológica.

Para la solución en la época de dominio de polvo se encuentra que $H_a = H_c$ y por lo tanto $\Theta = \frac{2}{3t}$, mientras que para el caso de la radiación se tiene que $\Theta = \frac{1}{2t}$ e igualmente se encuentra que $H_a = H_c$.

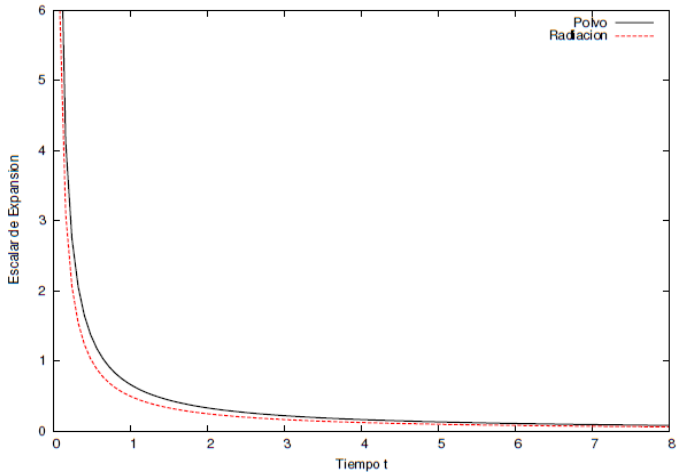

**Figura 3**. Evolución temporal del escalar de expansión

Los resultados hallados son los mismos que se encuentran con la métrica FLRW, esto debido a la proporcionalidad entre los factores de escala en cada solución.

En la Figura 3 se muestra la evolución temporal del escalar de expansión $\Theta$ en los casos de dominio de polvo y radiación. Se puede observar que en el caso de la radiación, la expansión disminuye más rápido que en el universo dominado por polvo, y, además, en los dos casos, la expansión tiende a cero para tiempos muy largos. Es decir, la expansión del universo tiene una mayor contribución del polvo que de la radiación en toda su evolución.

Por lo tanto, con la métrica propuesta y con las soluciones halladas se describe un universo axisimétrico en expansión.

## 6. ROTACIÓN DE UN FLUIDO IDEAL

Analizamos la rotación de un fluido en reposo en el caso axisimétrico que se plantea. Esto nos permite establecer si el universo modelado tiene movimiento de rotación con respecto a algún eje.

Dado que el tensor de rotación $\omega_{\mu\nu}$ de un fluido está dada por la expresión matemática en (26)

$$\omega_{\mu\nu} = \frac{1}{2}\left(u_{\mu,\nu} - u_{\nu,\mu}\right) \quad (26)$$

donde $u_\mu = (1, 0, 0, 0)$ es su cuadrivelocidad, para la métrica planteada en (7) encontramos que

$$\omega_{\mu\nu} = 0 \quad (27)$$

para cualquier contenido de materia. Es decir, en (27) se comprueba que un fluido ideal en reposo no rota mientras se expande en el universo descrito por la métrica en (7). Con esto, el universo que describimos es intrínsecamente no rotante.

## 7. PARÁMETRO *SHEAR*

El parámetro *shear* es un tensor que mide la tasa de distorsión del flujo de materia, es decir, es un parámetro que permite verificar si una expansión es anisótropa pues indica la tendencia de un fluido a cambiar de una forma esférica a una forma elipsoidal (Russell et al., 2014).

Sus componentes $\sigma_{\mu\nu}$ están dadas por (28)

$$\sigma_{\mu\nu} = \frac{1}{2}\left(u_{\mu,\nu} + u_{\nu,\mu}\right) - \Theta\left(g_{\mu\nu} - u_\mu u_\nu\right) \quad (28)$$

donde $\Theta$ es el escalar de expansión y $g_{\mu\nu}$ el tensor métrico.

Considerando un fluido en reposo con $u_\mu = (1,0,0,0)$ y considerando la métrica en (7), se tiene que las componentes no nulas del tensor *shear* son las indicadas en (29), (30), (31) y (33).

$$\sigma_{rr} = \frac{2}{3}c^2 \cos^2\phi(H_a - H_c) + \frac{1}{3}a^2 \sin^2\phi(H_c - H_a) \quad (29)$$

$$\sigma_{r\phi} = -r'\left[\frac{1}{3}a^2(H_a - H_c) + \frac{2}{3}c^2(H_a - H_c)\right]\cos\phi\sin\phi \quad (30)$$

$$\sigma_{\theta\theta} = -\frac{r'^2}{3}a^2(H_a - H_c)\sin^2\phi \quad (31)$$

$$\sigma_{\phi\phi} = -r'^2\left[\frac{2}{3}c^2 \sin^2\phi(H_c - H_a) + \frac{1}{3}a^2 \cos^2\phi(H_a - H_c)\right] \quad (32)$$

El escalar *shear*, que mide la tasa de distorsión en la expansión del fluido en alguna región (Russell et al., 2014), está dado por $\sigma^2 = \sigma_{ij}\sigma^{ij}$.

En el caso analizado, en función de las constantes de Hubble, se tiene que dicho escalar viene dado por

$$\sigma^2 = \frac{2}{3}(H_a - H_c)^2 \quad (33)$$

En (33) se puede notar que la no nulidad de las componentes del tensor y del escalar de *shear* depende de la diferencia entre los parámetros de Hubble.

Para el caso del universo vacío con constante cosmológica no nula, el universo lleno de polvo y el universo lleno de radiación, dado que $H_a = H_c$, se tiene que

$$\sigma_{\mu\nu} = 0 \quad (34)$$

Por lo tanto, a partir del resultado obtenido en (34), un fluido ideal en reposo se expande de forma isótropa en el universo descrito por (7). Para que la expansión pierda la forma esférica es necesario incluir fuentes de materia donde se cumpla que Ha ≠ Hc.

Dados estos resultados, nos preguntamos si es posible hallar una fuente de materia para la cual se cumpla que $H_a \neq H_c$, con lo cual, la expansión de un fluido ideal sea anisótropa. Para esto es necesario incluir fuentes anisótropas al modelo y estudiar las implicaciones en su cinemática.

Por otro lado, es importante también incluir en el modelo con fuentes de materia la constante cosmológica no nula para comprender sus efectos en la expansión anisótropa de un fluido ideal.

## 8. CONCLUSIONES

Se planteó una métrica axisimétrica anisótropa a partir de una transformación de coordenadas que describe un universo no rotante, en expansión y con un tensor de *shear* no diagonal con componentes no nulas.

Se establecieron tres ecuaciones de campo que deben resolverse y que son las equivalentes a las ecuaciones de Friedmann en el caso axisimétrico. Además, se planteó la ecuación de continuidad.

Se estudiaron tres casos: el vacío con constante cosmológica distinta de cero, la época de dominio del polvo y la época de dominio de la radiación. Para estos casos se estableció que las ecuaciones de campo se resuelven si los factores de escala son proporcionales entre sí.

Encontramos que la expansión anisótropa depende de la diferencia entre los parámetros de Hubble $H_a$ y $H_c$. Para los casos analizados se encontró que $H_a = H_c$, por lo tanto, un fluido ideal en reposo en el universo axisimétrico que planteamos se comporta como en un universo isótropo (tensor de *shear* nulo) en coordenadas comóviles.

## REFERENCIAS

Akarsu, O., Dereli, T., y Oflaz, N. (2014). Accelerating anisotropic cosmologies in Brans-Dicke gravity coupled to a mass-varying vector field. *Classical and Quantum Gravity*. 31. 045020

Amanullah, R. et al. (2010). Spectra and HST light curves of six type Ia Supernovae at 0.511<z<1.12 and the Union2 compilation. *Astrophys.J*. 716. 712-738

Cáceres, D. (n.d.). Los modelos cosmológicos homogéneos y anisotrópicos de Bianchi. *Congreso Colombiano de Astronomía y Astrofísica*. Obtenido de http://168.176.8.14/publicaciones/documentos/cocoa2008/cocoa15.pdf. (Junio, 2015).

Fernández, R. (2014). Implicaciones cosmológicas de las anisotropías de temperatura y polarización de la RFCM y la estructura a gran escala del universo (Tesis doctoral). Universidad de Cantabria, Cantabria, España.

Gourgoulhon, E. et al. (2015). SageManifolds Differential geometry and tensor calculus with Sage (version 0.8). http://sagemanifolds.obspm.fr
Jacobs, K. (1968). Bianchi Type I Cosmological Models (Tesis doctoral). California Institute of Technology, Pasadena, California, EE.UU.

Pankaj, J. (n.d.). Large scale anisotropy in the Universe. *I.I.T. Kanpur*. Obtenido de http://www.iitg.ernet.in/DAE-HEP2014/public/uploads/07-2-2014-54842f4b34efa0.77962936.pdf. (Junio, 2015).

Pavan, K., y Pankaj, J. (2012). Large scale anisotropy due to pre-inflationary phase of cosmic evolution. *Mod.Phys.Lett. A*. 27. 1250014

Pradhan, A., Saha, B., y Rikhvitsky, V. (2015). Bianchi type-I transit cosmological models with time dependent gravitational and cosmological constants-reexamined. *Indian Journal of Physics*.89. 503-513

Russell, E., Battan, C., y Pashaev, O. (2014). Bianchi I Model: An Alternative Way To Model The Presentday Universe. *MNRAS*. 442 (3). 2332-2341

Stein, W. A. et al. (2015). Sage Mathematics Software (Version 6.6). *The Sage Development Team*. http://www. sagemath.org